\documentclass[twocolumn]{aastex62}
\graphicspath{{./}{figures/}}

\received{June 21, 2023}
\accepted{September 8, 2023}
\submitjournal{ApJ}

\shortauthors{Pessi et al.}

\begin{document}

\title{A metallicity dependence on the occurrence of core-collapse supernovae} 

\correspondingauthor{Thallis Pessi}
\email{thallis.pessi@mail.udp.cl}

\author[0000-0001-6540-0767]{Thallis Pessi}
\affil{Instituto de Estudios Astrof\'isicos, Facultad de Ingenier\'ia y Ciencias, Universidad Diego Portales, Av. Ej\'ercito Libertador 441, Santiago, Chile}
\affil{European Southern Observatory, Alonso de Córdova 3107, Vitacura, Casilla 19001, Santiago, Chile}

\author[0000-0003-0227-3451]{Joseph P. Anderson}
\affil{European Southern Observatory, Alonso de Córdova 3107, Vitacura, Casilla 19001, Santiago, Chile}
\affiliation{Millennium Institute of Astrophysics MAS, Nuncio Monsenor Sotero Sanz 100, Off. 104, Providencia, Santiago, Chile}

\author[0000-0002-3464-0642]{Joseph D. Lyman}
\affil{Department of Physics, University of Warwick, Coventry CV4 7AL, UK}

\author[0000-0003-1072-2712]{Jose L. Prieto}
\affil{Instituto de Estudios Astrof\'isicos, Facultad de Ingenier\'ia y Ciencias, Universidad Diego Portales, Av. Ej\'ercito Libertador 441, Santiago, Chile}
\affiliation{Millennium Institute of Astrophysics MAS, Nuncio Monsenor Sotero Sanz 100, Off. 104, Providencia, Santiago, Chile}

\author[0000-0002-1296-6887]{Lluís Galbany}
\affil{Institute of Space Sciences (ICE, CSIC), Campus UAB, Carrer de Can Magrans, s/n, E-08193 Barcelona, Spain}
\affiliation{Institut d’Estudis Espacials de Catalunya (IEEC), E-08034 Barcelona, Spain}

\author[0000-0001-6017-2961]{Christopher S. Kochanek}
\affil{Department of Astronomy, The Ohio State University, 140 West 18th Avenue, Columbus, OH 43210, USA}
\affiliation{Center for Cosmology and Astroparticle Physics, The Ohio State University, 191 W. Woodruff Avenue, Columbus, OH 43210, USA}

\author[0000-0001-6444-9307]{Sebastian F. Sánchez}
\affil{Instituto de Astronomía, Universidad Nacional Autónoma de  México, A.~P. 70-264, C.P. 04510, México, Ciudad de México, Mexico}

\author[0000-0002-1132-1366]{Hanindyo Kuncarayakti}
\affiliation{Tuorla Observatory, Department of Physics and Astronomy, FI-20014 University of Turku, Finland}
\affil{Finnish Centre for Astronomy with ESO (FINCA), FI-20014 University of Turku, Finland}

\begin{abstract}

Core-collapse supernovae (CCSNe) are widely accepted to be caused by the explosive death of massive stars with initial masses $\gtrsim  8$~M$_\odot$. There is, however, a comparatively poor understanding of how properties of the progenitors -- mass, metallicity, multiplicity, rotation etc. --- manifest in the resultant CCSN population. 
Here we present a minimally biased sample of nearby CCSNe from the ASAS-SN survey whose host galaxies were observed with integral-field spectroscopy using MUSE at the VLT. 
This dataset allows us to analyze the explosion sites of CCSNe within the context of global star formation properties across the host galaxies.
We show that the CCSN explosion site oxygen abundance distribution is offset to lower values than the overall HII region abundance distribution within the host galaxies.
We further {split the sample at $12 + \textrm{log}_{10}(\textrm{O/H}) = 8.6$~dex and} show that within the subsample of low-metallicity host galaxies, the CCSNe unbiasedly trace the star-formation with respect to oxygen abundance, while for the sub-sample of higher-metallicity host galaxies, they preferentially occur in lower-abundance star-forming regions.
We estimate the occurrence of CCSNe as a function of oxygen abundance per unit star formation, and show that there is a strong decrease as abundance increases.
Such a strong and quantified metallicity dependence on CCSN production has not been shown before. Finally, we discuss possible explanations for our result and show that each of these has strong implications {not only for our understanding of CCSNe and massive star evolution, but also star-formation and galaxy evolution.}
 
\end{abstract}


\section{Introduction} \label{sec:intro}

Core-collapse supernovae (CCSNe) are generated by the explosive death of massive stars, and are important for the formation of new elements and the evolution of galaxies \citep{1995ApJS..101..181W,2008MNRAS.389.1137S,2019Sci...363..474J}.
Because the delay between star formation (SF) and death for massive stars is relatively short, CCSNe are expected to be associated with the SF regions of galaxies. 
Indeed, the connection between CCSNe and young and massive stars has been demonstrated by direct progenitor detections and by a large number of previous environment studies \citep[see, e.g.,][and references therein]{2015PASA...32...16S,2015PASA...32...19A}.
Massive stars produce strong ionising fluxes that excite the local ISM, producing HII regions. The latter are thus strong tracers of massive star formation within galaxies - specifically H$\alpha$ emission can be used as a direct tracer of the star formation rate (SFR) within galaxies \citep[e.g.,][]{1998ARA&A..36..189K}.

A CCSN explosion happens either by the collapse of an ONeMg core, for progenitor stars with initial masses $\approx 8-10$~M$_\odot$ \citep[also known as electron-capture SNe,][]{1980PASJ...32..303M, 1982Natur.299..803N}, or the collapse of an Fe core for $\gtrsim 10$~M$_\odot$ \citep{1979NuPhA.324..487B, 1990RvMP...62..801B}\footnote{Although the exact dividing initial mass remains uncertain.}. 
Theoretical models and simulations have demonstrated that the mass ranges of stars above these initial values that produce a CCSN explosion (i.e., that do not collapse directly to a black hole) is not continuous, with many regios of ``explodability'' \citep[see, e.g.,][]{2015ApJ...801...90P, 2016ApJ...821...38S, 2020ApJ...888...91E, 2021A&A...656L..19Z}. 

Stellar evolution models and CCSN simulations suggest that black hole formation should be more common at low metallicity \citep{2003ApJ...591..288H, 2004MNRAS.353...87E,2015ApJ...801...90P}, however, stellar population modelling suggests there is no strong dependence on the CCSN rate with metallicity \citep[e.g.,][]{2022MNRAS.514.1315B}.
Previous observational analyzes of CCSN production found only small \citep[e.g.,][]{2011MNRAS.412.1473L,2017ApJ...837..120G, 2021MNRAS.500.5142F} or no dependence  \citep[e.g.,][]{2021ApJS..255...29S} on metallicity (although these previous works used heterogeneous samples and metallicities derived from global galaxy properties).

In this work, we take advantage of a homogeneous CCSN host galaxy sample from the All-Sky Automated Survey for Supernovae (ASAS-SN) survey \citep{2014ApJ...788...48S, 2017PASP..129j4502K, 2017MNRAS.464.2672H, 2017MNRAS.467.1098H, 2017MNRAS.471.4966H, 2019MNRAS.484.1899H, 2023MNRAS.520.4356N} and the large field of view and fine spatial resolution of the {integral-field spectrograph} MUSE \citep[][]{2014Msngr.157...13B} to analyze how progenitor metallicity affects CCSN production.
Using this sample, in {\citet[][Paper I]{2023arXiv230611961P}}, we analyzed the differences in the local environment of the different CCSN types and in Pessi et al. (2023, in prep.; Paper II) we analyzed the CCSN environments in the context of the properties of all HII regions within their host galaxies.
Here, we present the results of comparing the local CCSN environment metallicity to the overall metallicity distributions of the SF regions in their host galaxies, and discuss the implications of the metallicity dependence on CCSN occurrence that we find. This letter is organized as follows: In Section \ref{sec:data} we present our sample and methods and in Section \ref{sec:res} we show the results of comparing the local CCSN environment to the overall SF across their host galaxies. In Section \ref{sec:disc} we discuss the implications of these results and in Section \ref{sec:conc} the conclusions are summarized.

\section{Data and Methods} \label{sec:data}

We analyze 98 galaxies that hosted 99 CCSNe detected by the ASAS-SN survey \citep{2017MNRAS.464.2672H, 2017MNRAS.467.1098H, 2017MNRAS.471.4966H, 2019MNRAS.484.1899H, 2023MNRAS.520.4356N} and were observed by MUSE, mostly in the context of the All-weather MUse Supernova Integral field Nearby Galaxies (AMUSING) survey \citep[][Galbany et al., in prep.]{2016MNRAS.455.4087G}. 
MUSE has a spatial sampling
of $0.2 \times 0.2 $~arcsec, a field of view of $1 $~arcmin$^2$ (in Wide Field Mode), and a mean spectroscopic resolution of $R \ {\sim} \ 3000$ with wide wavelength coverage ($480 - 930$~nm), enabling a detailed spectroscopic analysis of all HII regions within the observed galaxies.
The CCSN sample was detected by ASAS-SN between 2014 and 2018, and was reported in \citet{2017MNRAS.467.1098H, 2017MNRAS.471.4966H, 2019MNRAS.484.1899H} and  \citet{2023MNRAS.520.4356N}. 
Using a homogeneous sample from ASAS-SN minimizes biases in host properties because it is an untargeted and almost spectroscopically complete survey \citep[see e.g.,][for technical details on the survey]{2014ApJ...788...48S, 2017PASP..129j4502K}.
A description of the sample selection, data reduction and calibration is given in Paper I. 
From the 111 host galaxies in the initial AMUSING/ASAS-SN sample\footnote{\url{https://sites.google.com/view/theamusingasassnsample}}, we selected 78 that hosted Type II SNe (hereafter SNe~II) and 21 that hosted stripped-envelope SNe (SESNe, 7 SNe~IIb, 7 SNe~Ib, 4 SN~Ic, 2 SN~Ic-BL and 1 ambiguous SN~Ibc)\footnote{We remove all interacting events (IIn, Ibn) given the less clear nature of their progenitors.} 
The CCSN host galaxies are nearby, with luminosity distances of $10 \lesssim D_L \,({\rm Mpc}) \lesssim 169$.

We characterize the CCSN host galaxies through HII region segmentation, by combining the observed {H$\alpha$} signal of adjacent spaxels into one region.
We use \textsc{IFUanal}\footnote{\url{https://ifuanal.readthedocs.io/en/latest/index.html}} \citep{2018MNRAS.473.1359L} and the ``nearest bin'' method to perform the segmentation and obtain emission line and stellar continuum fits for each HII region in addition to that of the SN. 
Stellar continuum fitting is performed with \textsc{STARLIGHT} \citep[][]{2005MNRAS.358..363C, 2006MNRAS.370..721M} and multiple Gaussians are then fit to the emission lines of the continuum-subtracted spectra.

For each HII region (including that of each CCSN explosion site), the oxygen abundance (hereafter $[O/H]$ - a proxy for metallicity) is estimated following \citet[][D16]{2016Ap&SS.361...61D}, using the ratios of [{N}{II}]$\lambda 6584$ to [{S}{II}]$\lambda  \lambda  6717,31$ and [{N}{II}]$\lambda 6584$ to H$\alpha$. 
The D16 index is calibrated using photoionization models and is insensitive to reddening due to the nearby line ratios. It has been shown that D16 removes any dependence on the inferred abundance with ionization \citep{2017A&A...602A..85K}. In Appendix \ref{app:o3n2_n2}, we repeat part of our analysis using the commonly employed O3N2 and N2 diagnostics of \citet{2013A&A...559A.114M}. 
We also extract the {H$\alpha$} flux of each HII region and use the luminosity distance (derived from the CMB frame redshift), the host galaxy extinction correction from the Balmer decrement, and the Galactic reddening correction to estimate the H$\alpha$ luminosity (L$_{H\alpha}$) that is converted into a SFR as explained below.

{Although the lifetime of an HII region \citep[$\sim 10$~Myr, e.g.,][]{2014A&A...568A...4T, 2018MNRAS.477..904X} is shorter than the delay-time of CCSNe \citep[up to $\sim 200$~Myr, when considering binary evolution, e.g.,][]{2017A&A...601A..29Z, 2022MNRAS.514.1315B}, there should not be a significant abundance evolution in a couple generations of new SF. Therefore, the considered HII regions are a reasonable proxy for the progenitor abundance.}

Given that the amount of ionizing radiation per unit SF is a function of metallicity\footnote{{Other effects, such as $\alpha$-enhancement could affect the ionising flux as a function of oxygen abundance \textbf{\citep[e.g.,][although in this work only colder and older stellar atmospheres were althered, with the effect being uncertain for younger stellar populations]{2022MNRAS.512.5329B}.}}}, we use a Binary Population and Spectral Synthesis \citep[\textsc{BPASS},][]{2017PASA...34...58E, 2018MNRAS.479...75S} prescription to correct for this effect and obtain SFRs from HII region H$\alpha$ luminosities.
\textbf{We use a BPASS prescription with binary stellar populations, an initial mass function (IMF) consistent with \citet{1993MNRAS.262..545K} and an upper mass limit of $300$~M$_\odot$, as described in \citet{2017PASA...34...58E}.}

To characterize the star forming regions within the galaxies, we examine the cumulative distributions of SFR sorted by $[O/H]$ \citep[for a previous use of this method, see][]{2018MNRAS.473.1359L}.
For $i=1 \cdots N$ HII regions with SFR$_{i}$ and $[O/H]_i$ sorted by the oxygen abundance, we examine the integral distribution
\begin{equation}
    \label{eqn:1}
    \textrm{SFR}(<[O/H]_i) = \textrm{SFR}_{tot}^{-1} \sum_{j=1}^i \textrm{SFR}_{j}
\end{equation}
where $\textrm{SFR}_{tot}= \sum_i \textrm{SFR}_{i}$ is the total SFR of all the HII regions within our sample. 
{This gives the fraction of SF in each region compared to the total SF of the full sample as a function of metallicity.}

\section{Results} \label{sec:res}

\begin{figure}[t!]
\centering
\includegraphics[width=0.43\textwidth]{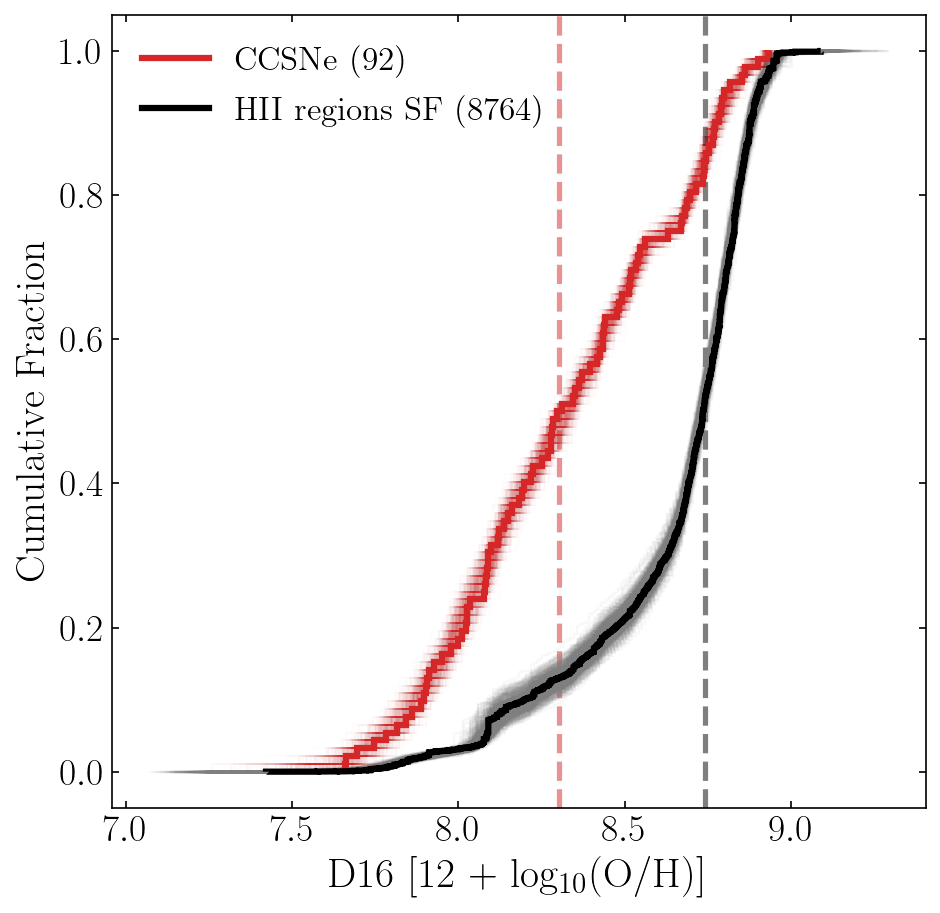}
\caption{The black line shows the cumulative distribution of SFR sorted by $[O/H]$ for all the HII regions in the galaxies, while the red line shows the cumulative distribution of $[O/H]$ at CCSN sites. 
The shaded regions show 500 bootstrap resamplings of each distribution (using the errors on the measured parameters) and the solid lines show the median of the realizations.
The black and red dashed lines show the median $[O/H]$ of the distributions for the HII regions and CCSNe, respectively.
\label{fig:CCSN_HII_weighted}}
\end{figure}

{Figure \ref{fig:CCSN_HII_weighted} shows the overall cumulative distribution of SFR sorted by $[O/H]$ for the CCSN environments (in red) and for all the 8810 identified HII regions within the galaxies of the sample (in black), as in Equation \ref{eqn:1}.}
The shaded regions show 500 bootstrap resamplings of each distribution and the solid lines show the median of the realizations.
The resampling is done by adding a Gaussian scatter on the measured errors of $[O/H]$ and SFR, {with the Gaussians centered at the $[O/H]$ and SFR values, using their uncertainties for the standard deviation}. 
The CCSNe have a median $[O/H]$ ${\sim} \ 8.30$~dex (red dashed line), while the overall SF has a median $[O/H]$ ${\sim} \ 8.74$~dex (black dashed line). 
This suggests that CCSN production does not follow the metallicity distribution of SF within all galaxies, but it is instead biased towards SF regions with lower abundance. 
\footnote{{If older HII regions had lower oxygen abundances then this could produce some of the effect we observe. However, we do not find a correlation between age, estimated by the H$\alpha$ EW, and oxygen abundance in our sample. These results will be outlined in Paper 2.}}

\begin{figure}[t!]
\centering
\includegraphics[width=0.46\textwidth]{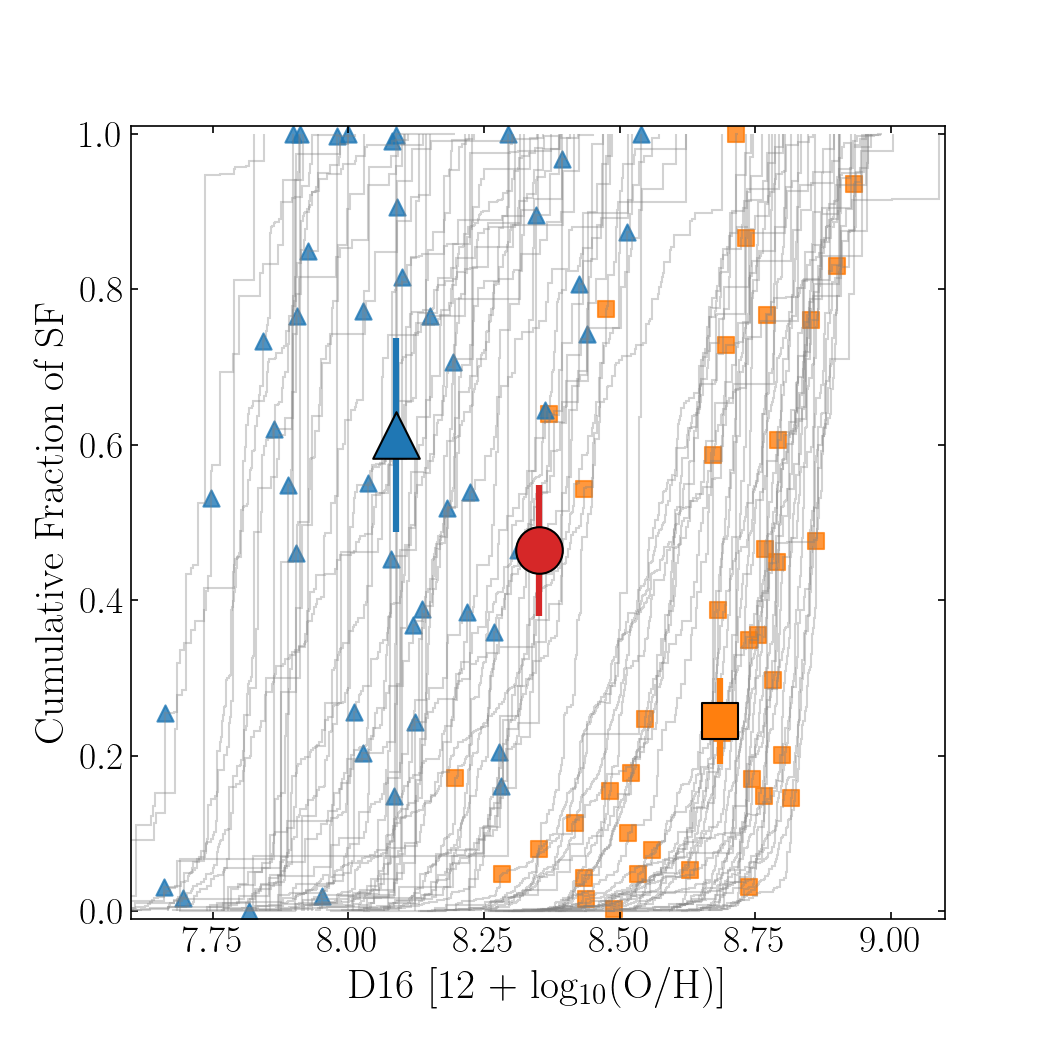}
\caption{Cumulative distributions of the fraction of SF as a function of $[O/H]$ for HII regions within individual galaxies.
Each gray line is the distribution for one galaxy and the small symbols are the SF fraction rank and $[O/H]$ of the CCSN environments within their host galaxy SF distributions. 
The median rank is given by the big markers, with uncertainties estimated by the median of the $1 \sigma$ of 500 bootstrap resamplings of each distribution. 
The blue triangles and orange squares represent the galaxies in which the HII region with the highest $[O/H]$ has a value lower or higher than $12 + \textrm{log}_{10}(\textrm{O/H}) = 8.6$~dex ({indicated by the dashed line}), respectively. 
The red circle shows the median values for the full sample.
\label{fig:cum_dist_hosts_ZD16_2}}
\end{figure}

\begin{figure}[t!]
\centering
\includegraphics[width=0.46\textwidth]{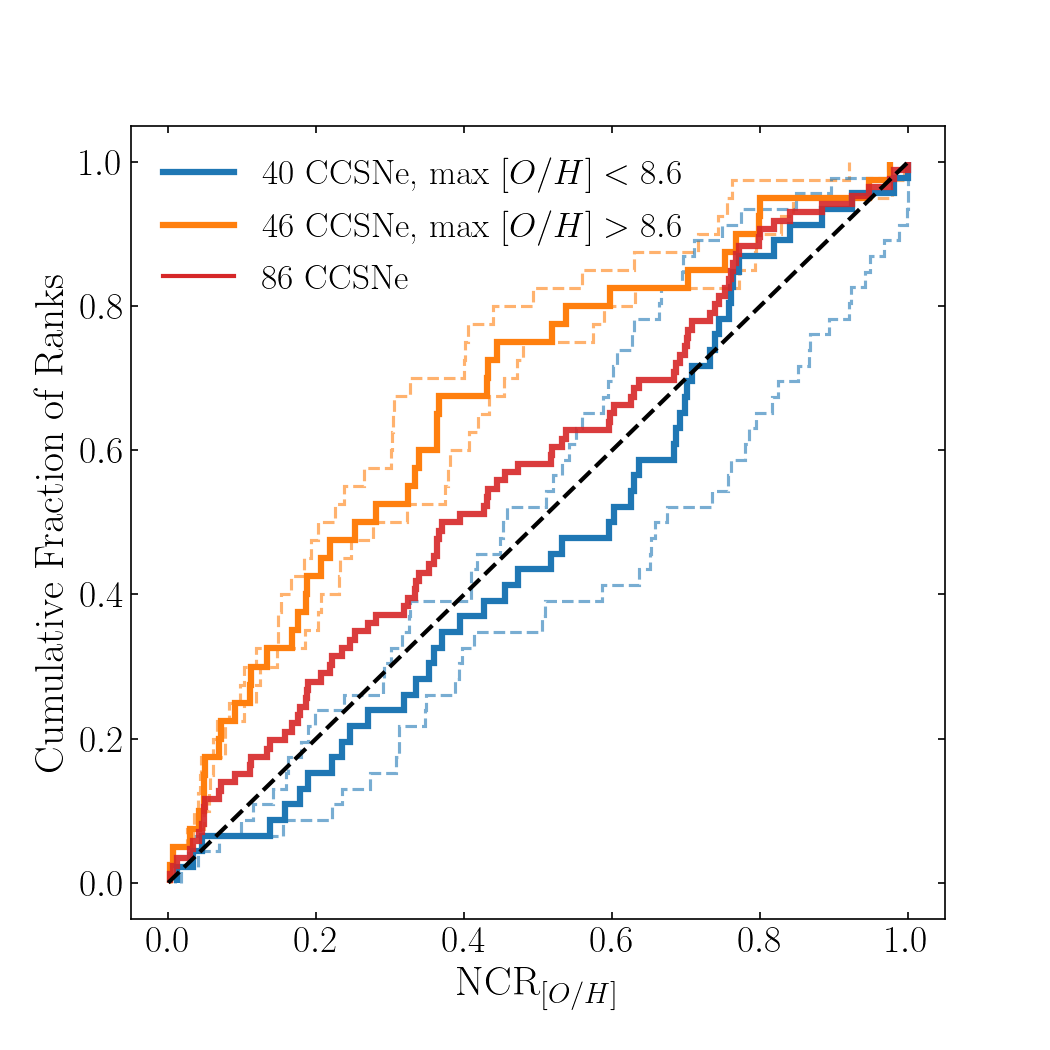}
\caption{The oxygen abundance NCR for the CCSNe. 
The red line shows the distribution for the full sample, and the blue and orange lines show, respectively, the galaxies where the largest metallicity value of their HII regions is lower or higher than $12 + \textrm{log}_{10}(\textrm{O/H}) = 8.6$~dex.
An unbiased distribution with respect to metallicity should follow the dashed diagonal line in the plot. 
The solid lines show the median of 500 bootstrap resamplings of each cumulative distribution and the dashed lines show the $1 \sigma$ of the realizations.
\label{fig:ZNCR}}
\end{figure}

The grey lines in Figure \ref{fig:cum_dist_hosts_ZD16_2} show the cumulative distributions of SF as a function of $[O/H]$ for all the HII regions within each of the individual {86 CCSN host galaxies from our initial sample}.
{From the initial sample (presented in Figure \ref{fig:CCSN_HII_weighted}), \textbf{we could not estimate} a L$_{H\alpha}$ for their associated HII region for six CCSNe (SN2018yo, SN2017gbv, ASASSN-17oj, ASASSN-16ba, ASASSN-16al, and SN2015W) due to the low signal-to-noise ratio, so they are excluded from the analyses presented in Figures \ref{fig:cum_dist_hosts_ZD16_2} and \ref{fig:ZNCR}. We also remove seven CCSNe (SN2015bm, ASASSN-15lv, ASASSN-16gy, ASASSN-18ou, SN2017ewx, SN2017ivu, SN2016afa) from the initial sample, where only one HII region in the host galaxy was identified.}
The small markers show the SF fraction rank { \citep[i.e., the $y-$axis position in the cumulative distribution, e.g.,][]{2016MNRAS.455.4087G, 2018ApJ...855..107G}} and $[O/H]$ of each CCSN environment within each distribution, and the large markers show the median of these values for the CCSNe. 
A uniform distribution with a median SF fraction of 0.5 would indicate that the SNe do not show any preference for occurring in regions with higher or lower values of $[O/H]$ within their host galaxies. 
The $[O/H]$ values range from $12 + \textrm{log}_{10}(\textrm{O/H}) \ {\sim} \ 7.4$~dex to ${\sim} \ 9.0$~dex, and the red circle shows the median $[O/H]$ of $12 + \textrm{log}_{10}(\textrm{O/H}) \ {\sim} \ 8.35$ for the full sample. The median SF fraction of $0.46 \pm 0.08$ suggests an unbiased production of CCSNe as a function of metallicity (but see below).
The uncertainties of the rank fractions are again estimated using bootstrap resampling.

Next, in Figure \ref{fig:cum_dist_hosts_ZD16_2}, we now split the galaxies at $12 + \textrm{log}_{10}(\textrm{O/H}) = 8.6$~dex \citep[nearly solar,][]{2021A&A...653A.141A} based on the HII region in the galaxy with the highest abundance. The blue triangles and orange squares show the SN fraction ranks of these low and high abundance galaxies, respectively. 
The median SF fraction of the CCSNe in the low metallicity galaxies is $0.61 \pm 0.12$, but the median SF fraction for the high metallicity galaxies is $0.24 \pm 0.05$. 
Those SNe within higher metallicity galaxies occur in relatively lower metallicity regions within their hosts, in contrast to those in the lower metallicity galaxies.

We analyze the two distributions of SF fraction ranks through the Normalised Cumulative Rank (NCR) of the CCSNe, shown in Figure \ref{fig:ZNCR}.
The NCR is based on the method presented in \citet{2006A&A...453...57J}, but taking the rank of each CCSN in the cumulative fractions of SF as a function of $[O/H]$ for their host galaxies \citep[see][for a similar application of this method]{2018MNRAS.473.1359L}.
The $x-$axis in Figure \ref{fig:ZNCR} becomes the SF fraction that is the $y-$axis in Figure \ref{fig:cum_dist_hosts_ZD16_2}.
A distribution of SNe that is unbiased with respect to abundance would follow the diagonal (i.e., a uniform distribution of ranks). A distribution biased to higher metallicities will lie below the diagonal and one biased to lower metallicities will lie above it.

The solid lines in Figure \ref{fig:ZNCR} show the median of the 500 bootstrap resamplings for the cumulative distributions, while the dashed lines show the $1 \sigma$ scatter realizations.
Figure \ref{fig:ZNCR} shows that for the full CCSN sample and the low metallicity galaxies, the SF fraction distribution follows the diagonal line, suggesting an unbiased dependence of these events with respect to metallicity. The Kolmogorov-Smirnov (KS) test \citep{chakravarti_1967} between the low metallicity galaxy distribution and a straight line has a $p-$value $= 0.69$, suggesting no significant metallicity bias. 
On the other hand, for higher metallicity galaxies the CCSN SF fractions are shifted to the left of the diagonal, with a KS-test $p-$value $= 0.03$. 
This suggests that they do not follow the distribution of SF as a function of metallicity in their host galaxies and are biased to SF at lower metallicities.

\begin{figure}[t!]
\centering
\includegraphics[width=0.45\textwidth]{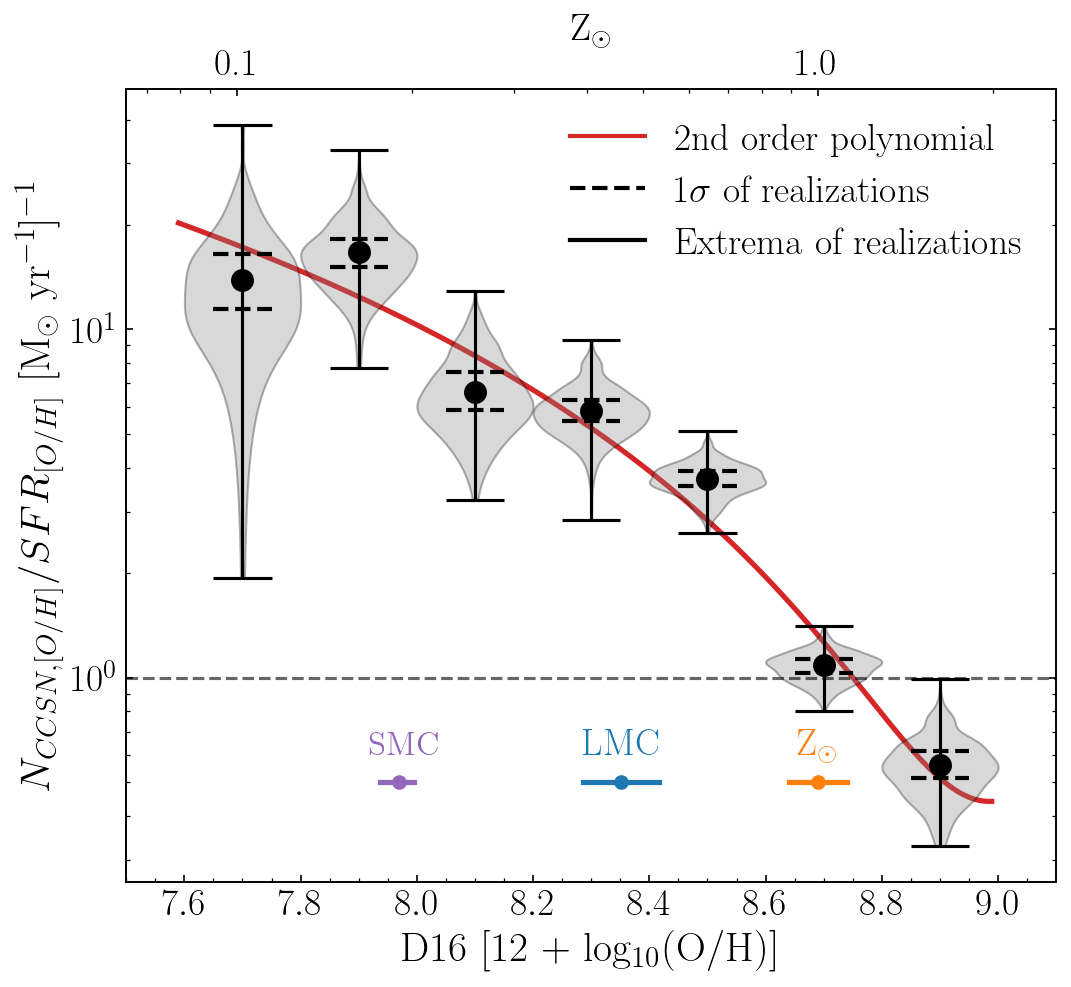}
\caption{The occurrence of CCSNe per unit of SF as a function of $[O/H]$. 
This is estimated by dividing the number of CCSNe, $N_{[O/H], CCSN}$, by the total SFR, $\textrm{SFR}_{tot, [O/H], i}$, over binned intervals in the abundance axis.
The violin plots show 500 bootstrap resamplings in each abundance interval of $0.2$~dex. 
The black dots show the median of these realizations, while the dashed and solid lines show, respectively, the $1 \sigma$ and the extrema of the realizations. 
We fit a 2nd order polynomial to the median values, weighting it by the $1 \sigma$ uncertainties.
{A horizontal dashed line is shown at $R_{[O/H], i} = 1$.}
The orange dot shows the abundance for the Sun \citep{2021A&A...653A.141A} and the purple and blue dots show the median abundances for, respectively, the SMC and LMC \citep{2017MNRAS.467.3759T}. 
\label{fig:falloff_function}}
\end{figure}

In Figure \ref{fig:falloff_function}, we divide the HII regions into seven 0.2~dex wide bins of abundance, and for each bin calculate
\begin{equation}\label{eq:2}
    R_{[O/H], i} = \frac{N_{CCSN, [O/H], i}}{\textrm{SFR}_{tot, [O/H], i}}
\end{equation}
where $N_{[O/H], CCSN, i}$ and $\textrm{SFR}_{tot, [O/H], i} = \sum_i \textrm{SFR}_{[O/H], i}$ are the numbers of CCSN and the total SFR (in units of M$_\odot$ yr$^{-1}$) associated with
abundance bin $i$. $R_{[O/H], i}$ is proportional to the occurrence rate of CCSN per unit of SF. Figure \ref{fig:falloff_function}
also shows ``violin'' errors from 500 bootstrap resamplings of the distributions. 
The median values of the resamplings, the 1$\sigma$ uncertainty, the central abundance and the number of SNe in each bin are reported in Table \ref{tab:falloff_function}.
Figure \ref{fig:falloff_function} also presents a second order polynomial fit of
\begin{equation}
    y = 0.11 - 2.25 \ x - 1.17 \ x^2
\end{equation}
between $12 + \textrm{log}_{10}(\textrm{O/H}) = 7.6$ and $9.0$~dex, where $y = \textrm{log}_{10} R_{[O/H]}$ and $x = [O/H] - 8.69$.

Figure \ref{fig:falloff_function} shows that the occurrence of CCSNe per unit of SF at lower oxygen abundance appears to be significantly larger than at higher abundances.
The occurrence rate declines as metallicity increases, with
$R_{[O/H]} \approx 15.5$ at $12 + \textrm{log}_{10}\textrm{(O/H)} \ {\sim} \ 7.7$~dex and $R_{[O/H]} \approx 0.5$ at $12 + \textrm{log}_{10}\textrm{(O/H)} \ {\sim} \ 8.9$~dex.
The latter value is nearly 25 times lower than the former, showing a large occurrence difference over metallicity.
Figure \ref{fig:falloff_function} also shows the abundances for the SMC, LMC \citep{2017MNRAS.467.3759T} and the Sun \citep{2021A&A...653A.141A}.
The occurrence of CCSNe per unit SF in abundances close to the SMC and LMC values are higher by factors of $\sim 15$ and $\sim 5$ than at abundances close to solar.
We show the occurrence of CCSN per unit of SF as a function of two other oxygen abundance indicators in Appendix \ref{app:o3n2_n2}, finding consistent results.

\begin{table}
\caption{The occurence rate of CCSN per unit of SF, as a function of abundance. \label{tab:falloff_function}}
\centering
\begin{tabular}{clcc}
\hline
N$_{SN}$ & $R([O/H])$ & $1 \sigma$ & $[O/H]$\\
\hline
7 & 15.54 & 4.94 & 7.70 \\
12 & 17.26 & 3.18 & 7.90 \\
16 & 7.01 & 1.65 & 8.10 \\
17 & 6.08 & 0.82 & 8.30 \\
20 & 4.00 & 0.37 & 8.50 \\
16 & 1.09 & 0.10 & 8.70 \\
8 & 0.50 & 0.10 & 8.90 \\
\hline
\end{tabular}
\end{table}

\section{Discussion} \label{sec:disc}

We have found that the CCSN production appears to be strongly dependent on metallicity, with the detected number of CCSNe relative to SF decreasing as oxygen abundances increase.
CCSNe do not follow the general distribution of SF as a function of metallicity and tend to happen within relatively lower metallicity regions in higher metallicity galaxies. In lower metallicity galaxies, the occurrence of CCSNe seems to be unbiased with respect to metallicity. 
When binning SNe and SF by oxygen abundance, we find a clear decreasing occurrence of CCSNe per unit SF with increasing [O/H].

Previous works have explored the metallicity dependence of the occurrence of long gamma-ray bursts \citep[LGRBs, which are connected to SNe Ic-BL,][]{2006AcA....56..333S,2006Natur.441..463F, 2007ApJ...654L..21S, 2007AJ....133..882K, 2008AJ....135.1136M, 2011MNRAS.414.1263M, 2013ApJ...774..119G, 2017ApJ...834..170G} and superluminous (SL)SNe \citep{2011ApJ...730...34S, 2013ApJ...763L..28C, 2013ApJ...771...97L, 2017MNRAS.470.3566C, 2021MNRAS.500.5142F}.
Some works also discussed a metalliticy effect on the rate of SNe Ia \citep{2006ApJ...648..868S, 2011MNRAS.412.1473L, 2013MNRAS.430.1746G, 2013ApJ...770...88K, 2017ApJ...837..120G, 2019MNRAS.484.3785B, 2022arXiv221001818J}. 
There are a number of studies of the correlations between CCSN rates and host properties.  In particular, \citet{2011MNRAS.412.1473L} found that the CCSN rates per unit luminosity (either $B$ or $K$ band) or stellar mass decreased with galaxy luminosity or mass using the LOSS SN sample \citep{2000AIPC..522..103L, 2001ASPC..246..121F}. Indeed,
the rates dropped by a factor of ${\sim} \ 5$ between host galaxy stellar masses of $10^9$~M$_\odot$ and $2 \times 10^{10}$~M$_\odot$. \citet{2017ApJ...837..120G} reanalyzed those data and found
changes in the CCSN rate per unit stellar mass of a factor of ${\sim} \ 6$ for Type II SN and a factor for ${\sim} \ 4$ for SESNe between galaxies oxygen abundances of ${\sim} \ 8.7$ to ${\sim} \ 9.2$~dex, which is similar to what we see in Figure \ref{fig:falloff_function}.
\citet{2017ApJ...837..120G} argue, however, that this is due to changes in the specific galaxy star formation rates rather than metallicity.  
\citet{2021MNRAS.500.5142F} also found a drop in the CCSN rates per unit stellar mass of a factor of ${\sim} \ 6$ between masses of
$10^9$ and $10^{11}$~M$_\odot$.
Here, we are directly examining the dependence of the occurrence of CCSN per unit SF as a function of metallicity, and the driver appears to be environment metallicity.

Our study has some significant advantages compared with previous analyzes: (1) we use a homogeneous sample of CCSNe detected by ASAS-SN and that (2) were selected in a minimally biased way to be observed by MUSE (see Paper I for a more detailed description of sample selection); and (3) we use emission-line spectroscopy of spatially resolved HII regions to measure both metallicities and SFRs across the galaxies (in contrast to central metallicities, taken, e.g., from SDSS, or global galaxy properties derived from photometry).

We now outline possible explanations for the result presented above. It is important to note that while the exact explanation of our result is currently unclear, any of the below possibilities -- if confirmed -- would have strong implications for our understanding of the CCSN phenomenon and/or massive SF within galaxies.

    The first possibility that we discuss is that our findings are the result of selection effects and that we are missing CCSNe within higher metallicity HII regions. This could be caused by higher extinction in higher metallicity regions, or the difficulty in detecting SNe in high surface-brightness regions (which are generally of higher metallicity). 
    While we do find that higher abundance HII regions have slightly higher $E(B-V)$ values than lower regions, this difference is not significant, neither is the overall distribution of host HII region extinction (see Paper I and Paper II). In addition, near-infrared surveys suggest that only around $20\%$ of CCSNe are likely to be missed in the local Universe due to extinction in optical surveys -- this is much less than required to explain our result \citep[e.g.][]{2021MNRAS.506.4199F}.
    It has also been demonstrated that ASAS-SN has a high efficiency in detecting (and classifying) SNe towards the centers -- and thus the most extinction affected regions -- of galaxies \citep{2014ApJ...788...48S, 2017PASP..129j4502K, 2017MNRAS.464.2672H, 2017MNRAS.467.1098H, 2017MNRAS.471.4966H, 2019MNRAS.484.1899H, 2023MNRAS.520.4356N}.
    ASAS-SN also found more tidal disruption events (TDEs) per SN than previous surveys, indicating a high efficiency for finding transients close or at the centers of galaxies \citep{2017MNRAS.464.2672H, 2017MNRAS.467.1098H, 2017MNRAS.471.4966H, 2019MNRAS.484.1899H}.
    Finally, it should be noted, any extinction effect that biases our CCSN distribution must also affect our HII SF distribution. If SNe are being missed due to high extinction, the underlying HII regions could also be missed, thus lessening the impact of extinction in our comparative analysis.

    Given that ASAS-SN is a magnitude limited survey, the number of discovered events scales with their peak luminosity.
    We have made no completeness correction here and essentially we are assuming that the CCSN luminosity function is not strongly abundance dependent. 
    If the CCSN peak luminosity were dependent on metallicity, this could affect our results. However, we have shown in Paper I that there is no strong correlation between peak luminosity and oxygen abundance at the location of CCSNe, making this explanation less compelling. 
        
    Thus, it seems unlikely that selection effects can explain our results. However, should such selection effects -- together with others not considered -- be responsible for our results, then this would still be a significant result. Indeed, to reproduce our observed number of CCSNe per unit SF as a function of metallicity (if no intrinsic metallicity difference exists) would require that we are missing more than a factor of 10 CCSNe at solar metallicity (per unit SF) compared to the SMC and lower metallicities {(as we show in Appendix \ref{app:o3n2_n2}, the result change little if we use difference abundance estimates)}. This would mean that currently estimated CCSN rates are hugely underestimated, with a significant impact on our understanding of massive star evolution and death.

    A second possibility is that there is a strong metallicity dependence to the IMF. Our analysis assumes that the IMF does not vary with environmental properties such as metallicity. This, however, could be the case: at low metallicity the shape of the IMF could be such that many more massive stars that will explode as CCSNe are produced per unit SF, while at high metallicity the IMF may produce many less CCSN progenitors.
    This scenario was explored by recent studies on the metallicity dependence of the IMF \citep[e.g.,][although any claimed dependence is much smaller than what would be required here]{2023Natur.613..460L}.
    Such a result would also be significant, and have many implications on our current understanding of star formation and galaxy evolution.

    Our third possible explanation is that H$\alpha$ is a significantly biased tracer of massive SF; as explained in Section \ref{sec:data}, we used a standard prescription for converting the H$\alpha$ luminosity into an estimate of the SFR which gives less SFR per unit H$\alpha$ luminosity at low metallicity because the model stellar populations produce more ionizing radiation per unit SF. If this dependence could be made weaker or even be reversed, this would help to explain our result. However, this would imply that the real metallicity dependence on the H$\alpha$ to SFR conversion is much higher than commonly used prescriptions.

    Finally, it could be that metallicity is a significant factor regulating the `explodability' of massive stars. This would imply that there are many more massive stars exploding as CCSNe within lower metallicity environments than at higher metallicity (e.g. solar abundance). 
    This result would have a significant impact in our understanding of stellar evolution and the explosion mechanism of massive stars, and the numbers of different types of compact remnants.
    This would go against stellar evolution studies and models that suggest that black hole formation is more common at lower progenitor metallicities {\citep[e.g.,][]{2003ApJ...591..288H, 2004MNRAS.353...87E, 2011ApJ...730...70O,2016ApJ...818..124E}, although the metallicity dependence on the expected outcome of failed SNe has yet not been explored in depth \citep[e.g.,][]{2008ApJ...684.1336K, 2013ApJ...768L..14P}.}
    Some studies suggest that the lower mass limit for CCSN decreases as a function of metallicity \citep[for those metallicity ranges probed in the current study; e.g.,][]{2004MNRAS.353...87E, 2013ApJ...765L..43I}, and that metallicity can also affect mass-loss and the populations of different SNe \citep[for a recent analysis see, e.g.,][]{2022A&A...661A..60A, 2023A&A...671A.134A}. 
    The effect of metallicity could be particularly stronger in binary systems. As demonstrated by \citet{2020A&A...637A...6L}, stripped stars in binaries at lower metallicity could have a higher chance of producing a SN, and \citet{2021A&A...656A..58L} shows that the compactness of binary-stripped stars is lower than single stars, making their explodability higher 
    {\citep[for previous studies on the influence of binary systems on the core structure, see e.g.,][]{1989A&A...220..135L, 1993ApJ...411..823W, 2021ApJ...916L...5V}}. 
    {Additionally, \citet{2020A&A...638A..55K} shows that envelope stripping is delayed at lower metallicities, resulting in a higher final core mass, although \citet[][]{2020A&A...634A..79S} demonstrate that binary interactions might not dominate the formation of WRs in low metallicity environments.}
    This, and the fact that stellar multiplicity might be higher at lower metallicity \citep[which has been shown to be the case for low-mass stars,][]{2018ApJ...854..147B}, could be at play in producing the effect we see in this work {\citep[although no strong evidence exists in the case of massive stars, see e.g.,][]{2013A&A...550A.107S, 2015A&A...580A..93D}.}

\section{Conclusions} \label{sec:conc}

We have demonstrated that the observed occurrence of CCSNe shows a strong dependence on environment metallicity. Selection effects would not seem to be able to explain such an effect; if they do, this has strong implications for the estimated rates of CCSNe. 
Our results also challenge the standard assumptions of a universal IMF and the use of H$\alpha$ as a tracer of SFR; a deep analysis of how these quantities are affected by local SF properties should be further explored.
Finally, our results could be explained by a strong dependence of metallicity on the explosion mechanism of massive stars; this effect has not been addressed in depth in the literature, and should be taken into account in future theoretical analyses.

\begin{acknowledgments}
We thank Elizabeth Stanway and Andrew Levan for the fruitful discussion that aided in our understanding of our results. We also thank Jan Eldridge for comments. 
Based on observations collected at the European Southern Observatory under ESO programme(s): 
      096.D-0296 (A), 
      0103.D-0440 (A),
      096.D-0263 (A), 
         097.B-0165 (A),
         097.D-0408 (A),
        0104.D-0503 (A),
         60.A-9301 (A),
         096.D-0786 (A),
         097.D-1054 (B),
         0101.C-0329 (D),
         0100.D-0341 (A),
         1100.B-0651 (A),
         094.B-0298 (A),
         097.B-0640 (A),
         0101.D-0748 (A),
         095.D-0172 (A),
         1100.B-0651 (A),
         0100.D-0649 (F),
         096.B-0309 (A).
T.P. acknowledges the support by ANID through the Beca Doctorado Nacional 202221222222.
    J.L.P. acknowledges support by ANID through the Fondecyt regular grant 1191038 and through the Millennium Science Initiative grant ICN12\_009, awarded to The Millennium Institute of Astrophysics, MAS. 
    L.G. acknowledges financial support from the Spanish Ministerio de Ciencia e Innovaci\'on (MCIN), the Agencia Estatal de Investigaci\'on (AEI) 10.13039/501100011033, and the European Social Fund (ESF) "Investing in your future" under the 2019 Ram\'on y Cajal program RYC2019-027683-I and the PID2020-115253GA-I00 HOSTFLOWS project, from Centro Superior de Investigaciones Cient\'ificas (CSIC) under the PIE project 20215AT016, and the program Unidad de Excelencia Mar\'ia de Maeztu CEX2020-001058-M.
  J.D.L. acknowledges support from a UK Research and Innovation Fellowship(MR/T020784/1).
    H.K. was funded by the Academy of Finland projects 324504 and 328898.
         
\end{acknowledgments}

\vspace{5mm}
\facilities{VLT (MUSE)}

\software{
\textsc{BPASS} \citep{2017PASA...34...58E, 2018MNRAS.479...75S},
ESO reduction pipeline \citep{2014ASPC..485..451W}, \textsc{EsoReflex} \citep{2013A&A...559A..96F}, \textsc{CosmicFlows} \citep{2015MNRAS.450..317C}, \textsc{Astropy} \citep{astropy:2013}, \textsc{SciPy} \citep{2020SciPy-NMeth}, \textsc{VizieR Queries} \citep{2019AJ....157...98G} and \textsc{IFUanal} \citep{2018MNRAS.473.1359L}.   
          }

\bibliography{ref}

\appendix

\section{O3N2 and N2 Indicators}\label{app:o3n2_n2}

\begin{figure}[t!]
\centering
\includegraphics[width=0.45\textwidth]{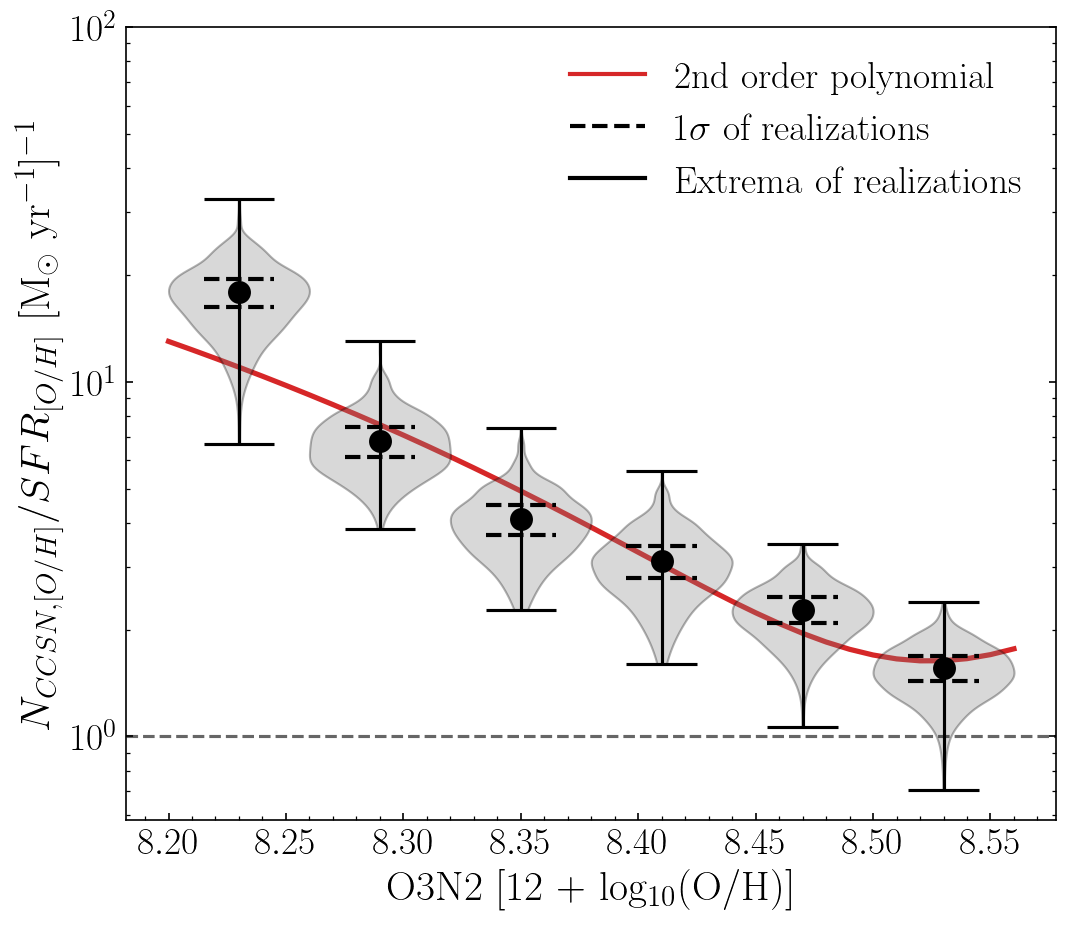}
\includegraphics[width=0.45\textwidth]{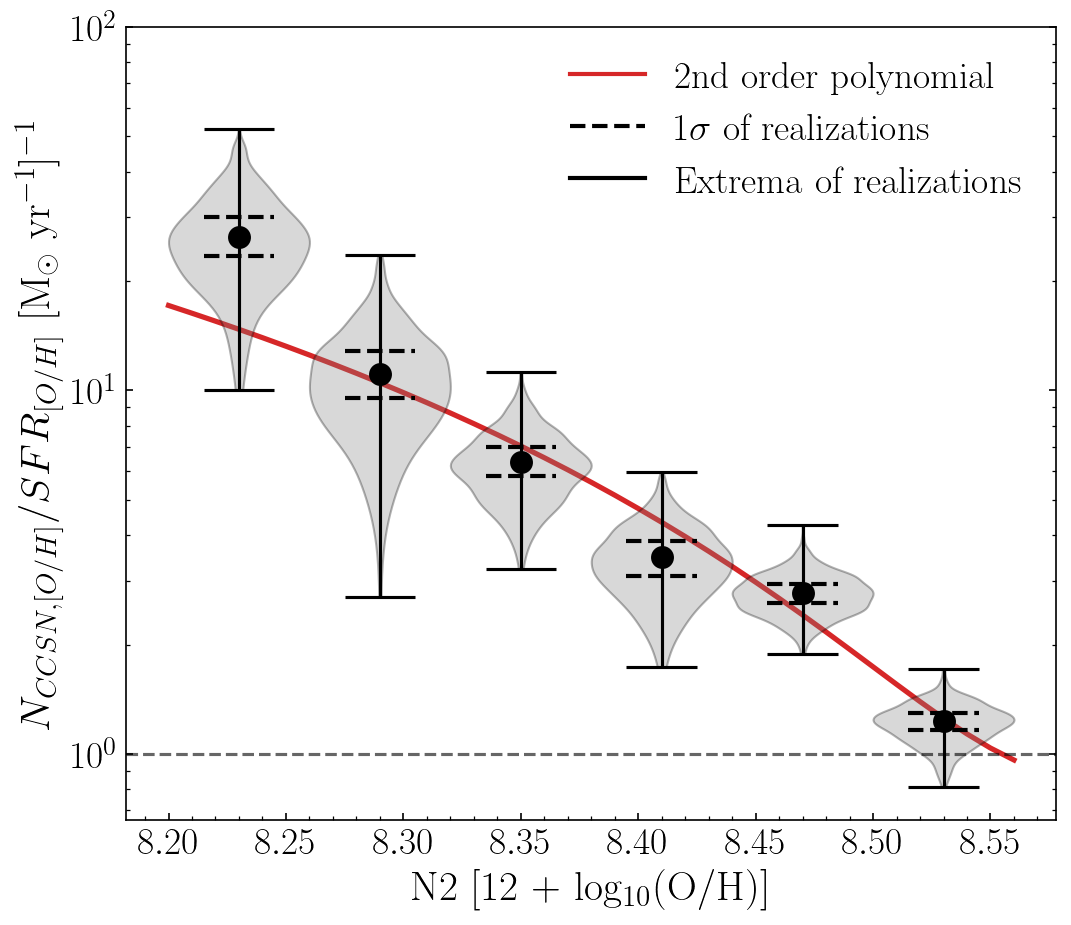}
\caption{The occurrence rate of CCSNe per unit of SF as a function of $[O/H]$, given by the O3N2 (left panel) and N2 (right panel) indicators. 
The violin plots show 500 bootstrap resamplings in abundance intervals of $0.06$~dex. 
The black dots show the median of these realizations, and the dashed and solid lines show, respectively, the $1 \sigma$ and the extrema of the realizations. 
A 2nd order polynomial is fit to the median values of each indicator, weighting it by the $1 \sigma$ uncertainties.
\label{fig:falloff_function_other_index}}
\end{figure}

The [{S}{II}] line used in the D16 index may be affected by ionization due to supernova remnants in galaxies, making it susceptible to contamination \citep[e.g.,][]{2020MNRAS.491..889K, 2021MNRAS.502.1386C}. 
We therefore estimate the occurrence of CCSN per unit of SF, given by Equation \ref{eq:2}, using indicators independent of [{S}{II}].
Figure \ref{fig:falloff_function_other_index} shows the resultant occurrence of CCSN per unit of SF as a function of oxygen abundance given by the O3N2 and N2 indexes, both estimated using the \citet[][]{2013A&A...559A.114M} calibration.
The O3N2 index uses the H$\alpha$, H$\beta$, [{O}{III}]$\lambda 5007$ and [{N}{II}]$\lambda 6584$ emission lines, while the N2 index uses the line ratio between [{N}{II}]$\lambda 6584$ and H$\alpha$.

Figure \ref{fig:falloff_function_other_index} shows a similar behavior for the occurrence of CCSNe per unit of SF as shown in Figure \ref{fig:falloff_function}, declining as metallicity increases. 
The $[O/H]$ values range between $8.2$ and $8.56$~dex, and we divide the HII regions between six $0.06$~dex bins of abundance.
In a similar way to Figure \ref{fig:falloff_function}, we show ``violin'' uncertainties from 500 bootstrap resamplings of the distributions, the 1$\sigma$ and extrema of the realizations given by, respectively, the dashed and solid lines.
We also show a 2nd order polynomial fit to the two indicators. 
The O3N2 index is fit with $y = 0.5 - 3.0 \ x + 4.0 \ x^2$, and the N2 index with $y = 0.7 - 3.9 \ x - 4.4 \ x^2$, between $12 + \textrm{log}_{10}(\textrm{O/H}) = 8.2$ and $8.56$~dex, where $y =  \textrm{log}_{10} R_{[O/H]}$ and $x = [O/H] - 8.4$. 
The O3N2 index has $R_{[O/H]} \approx 18$ at $12 + \textrm{log}_{10}\textrm{(O/H)} \ {\sim} \ 8.2$~dex and $R_{[O/H]} \approx 1.5$ at $12 + \textrm{log}_{10}\textrm{(O/H)} \ {\sim} \ 8.5$~dex, and the N2 index has $R_{[O/H]} \approx 25$ at $12 + \textrm{log}_{10}\textrm{(O/H)} \ {\sim} \ 8.2$~dex and $R_{[O/H]} \approx 1.2$ at $12 + \textrm{log}_{10}\textrm{(O/H)} \ {\sim} \ 8.5$~dex.
{There are large systematic uncertainties between the different abundance indicators \citep{2008ApJ...681.1183K}, which can lead to the quantitative differences seen here. However, the large difference factors seen in the three indicators suggest that the metallicity effect is not affected by these uncertainties.}
This indicates that the metallicity dependence on CCSN occurrence is independent of the chosen oxygen abundance indicator, being also observed to a similar degree when using O3N2 and N2.

\end{document}